\documentclass[11pt]{article}
\usepackage{moriond,epsfig,amssymb}

\newcommand{\Eq}[1]{Eq.\,(\ref{#1})}

\newcommand{\Fig}[1]{Figure \ref{#1}}

\newcommand{\beq}{\begin{equation}}                
\newcommand{\eeq}{\end{equation}}        
\newcommand{\bea}{\begin{eqnarray}}               
\newcommand{\eea}{\end{eqnarray}}        
\newcommand{\bdm}{\begin{displaymath}}                 
\newcommand{\edm}{\end{displaymath}}                      
  
\newcommand{\non}{\nonumber}
\newcommand{\bm}{\boldmath}

\newcommand{\ord}{{O}}

\newcommand{\mc}{m_c}
                
\newcommand{\mb}{m_b}   
   
\newcommand{\mt}{m_t}
\newcommand{\mll}{m_{\ell \ell}}  

\newcommand{\MW}{M_{\scriptscriptstyle W}}

\newcommand{\GeV}{\, {\rm GeV}}

\newcommand{\gs }{ g }
\newcommand{\as }{ \alpha_s }         
\newcommand{\aem}{ \alpha }   
\newcommand{\f}{\frac}

\newcommand{\Leff}{{\cal L}_{\rm eff}}

\newcommand{\LQCDQED}{{\cal L}_{{\rm QCD} \times {\rm QED}}}

\newcommand{\GF}{G_\mu}

\newcommand{\btosgamma}{b \to s \gamma}

\newcommand{\btoslpluslminus}{b \to s \ell^+ \ell^-}
\newcommand{\BtoXsgamma}{\bar{B} \to X_s \gamma}
\newcommand{\BtoXslpluslminus}{\bar{B} \to X_s \ell^+ \ell^-}

\newcommand{\btos}{b \to s}

\newcommand{\BtoKKstarlpluslminus}{\bar{B} \to K^{(\ast)} \ell^+ \ell^-}

\newcommand{\BtoXclnu}{\bar{B} \to X_c \ell \nu}
\newcommand{\BtoXulnu}{\bar{B} \to X_u \ell \nu}
\newcommand{\BR}{{\rm BR}}
\newcommand{\BRll}{{{\rm BR}_{\ell \ell}}}
\newcommand{\BRllexp}{{{\rm BR}_{\ell \ell}^{\rm exp}}}

\newcommand{\tinyskip}{\hspace{0.5mm}}
\newcommand{\Tinyskip}{\hspace{1.0mm}}

\begin{document}
\rightline{FERMILAB-Conf-04/063-T}

\vspace*{4cm}

\title{RECENT THEORETICAL DEVELOPMENTS IN \bm $\BtoXslpluslminus$
DECAYS\Tinyskip\footnote{Talk given at XXIXth Rencontres de Moriond
Electroweak Interactions and Unified Theories, La Thuile, 
Aosta Valley, Italy, March 21--28, 2004.}}     

\author{ U. HAISCH }

\address{Theoretical Physics Department, Fermilab, \\ Batavia, IL
60510, USA} 

\maketitle\abstracts{
We present a concise review of the theoretical status of the rare
semileptonic $\BtoXslpluslminus$ decays in the standard
model. Particular attention is thereby devoted to the recent
theoretical progress concerning, on the one hand the
next-to-next-to-leading order QCD calculation and, on the other hand
the analysis of phenomenological important subleading electroweak
effects.   
}   

\section{Introduction}
\label{sec:introduction}

The rare semileptonic $\btoslpluslminus$ transitions have been
observed for the first time by Belle and BaBar in form of the
exclusive $\BtoKKstarlpluslminus$
modes.\tinyskip\cite{exclusive,iwasaki,paolini} Recently also
inclusive measurements have become
available.\tinyskip\cite{iwasaki,paolini,inclusive} Like all other 
flavor-changing-neutral-current processes, these channels are
important probes of short-distance physics. Their study can yield
useful complementary information, when confronted with the less rare
$\btosgamma$ decays, in testing the flavor sector of the Standard
Model (SM). In particular, a precise measurement of the inclusive
$\BtoXslpluslminus$ mode would be welcome in view of New Physics (NP)
searches, because it is amenable to a clean theoretical description
for dilepton invariant masses, $\mll^2 \equiv q^2$, below and above
the $c \bar c$ resonances, namely in the ranges $1 \GeV^2
\le q^2 \le 6 \GeV^2$ and $q^2 \ge 14.4
\GeV^2$.\tinyskip\cite{Ghinculov:2003qd} Each window has its own
assets and drawbacks, related on the one hand to experimental issues,
such as event rates, identification and detection efficiencies, on
the other hand to theoretical questions, concerning the significance
of parametric errors, perturbative and non-perturbative effects. In   
the following we present results that will cover both regions,
performing a thorough study of the associated
uncertainties. Unfortunately, however, it is not possible to compare 
these predictions with experiment, as it is still to early to fit a
$q^2$ distribution to the data. To allow the comparison with the
existing inclusive measurements, we thus reexamine the SM prediction 
for the $\BtoXslpluslminus$ branching ratio, $\BRll$, including all
known QCD and electroweak effects. Finally, we update the SM result
for the position of the zero of the Forward-Backward (FB) asymmetry,
$q_0^2$.     

\section{Theoretical Framework}
\label{sec:framework}

The calculation of the partonic decay rate of $\BtoXslpluslminus$
consists of several parts that are worth recalling. Perturbative QCD
effects play an important role, due to the presence of large
logarithms of the form $L \equiv \ln \mb/\MW$, that can be resummed
using the machinery of Operator Product Expansion (OPE) and
Renormalization Group (RG) improved perturbation theory. Factoring  
out the Fermi constant $\GF$ and the electromagnetic coupling $\aem$,
the amplitude receives contributions of $\ord(\as^n L^{n + 1})$ at the
Leading  Order (LO), of $\ord( \as^n L^n)$ at the Next-to-Leading
Order (NLO), and of $\ord(\as^n L^{n-1})$ at the
Next-to-Next-to-Leading Order (NNLO) in QCD. 

To achieve the necessary resummation, one works in the framework of an
effective low-energy theory with five active quarks, three active
leptons, photons and gluons, obtained by integrating out heavy degrees
of freedom characterized by a mass scale $M \ge \MW$. At LO in the OPE
the effective on-shell Lagrangian relevant for the $\btoslpluslminus$
transition at a scale $\mu$ is given by  
\beq \label{eq:effectivelagrangian}
\Leff = \LQCDQED + \f{4 \GF}{\sqrt{2}} V^\ast_{ts} V_{tb} \sum_{i} C_i
(\mu) \, Q_i \, .      
\eeq
Here the first term is the conventional QCD--QED Lagrangian for the
light SM particles. In the second term $V_{ij}$ denotes the elements
of the Cabibbo-Kobayashi-Maskawa matrix and $C_i (\mu)$ are the Wilson
coefficients of the corresponding operators $Q_i$ built out of the
light fields.   

Neglecting for the moment subleading electroweak effects, as well as
the QCD penguin operators $Q_3$--$Q_6$, that are suppressed by small 
Wilson coefficients, the remaining physical operators arising in the
SM can be written as  
\beq \label{eq:physicaloperators} 
\begin{array}{lll}
Q_1 = ( \bar{s}_{\scalebox{0.65}{$L$}} \gamma_\mu T^a
c_{\scalebox{0.65}{$L$}} ) ( \bar{c}_{\scalebox{0.65}{$L$}} \gamma^\mu
T^a b_{\scalebox{0.65}{$L$}} ) \, , & & Q_2 = (
\bar{s}_{\scalebox{0.65}{$L$}} \gamma_\mu c_{\scalebox{0.65}{$L$}} ) (
\bar{c}_{\scalebox{0.65}{$L$}} \gamma^\mu b_{\scalebox{0.65}{$L$}} )
\, , \non \\[2mm] 
Q_7^\gamma = \scalebox{1.0}{$\f{e}{\gs^2}$} \mb (
\bar{s}_{\scalebox{0.65}{$L$}} \sigma^{\mu \nu}
b_{\scalebox{0.65}{$R$}} ) F_{\mu \nu} \, , & & Q_8^g =
\scalebox{1.0}{$\f{1}{\gs}$} \mb ( \bar{s}_{\scalebox{0.65}{$L$}} 
\sigma^{\mu \nu} T^a b_{\scalebox{0.65}{$R$}} ) G_{\mu \nu}^a \, ,
\\[2mm]            
Q_9 = \scalebox{1.0}{$\f{e^2}{\gs^2}$} (
\bar{s}_{\scalebox{0.65}{$L$}} \gamma_\mu b_{\scalebox{0.65}{$L$}} ) 
\scalebox{1.0}{$\sum\nolimits_\ell$} ( \bar{\ell} \gamma^\mu \ell ) \,
, & & \hspace{-1.5mm} Q_{10} = \scalebox{1.0}{$\f{e^2}{\gs^2}$} ( 
\bar{s}_{\scalebox{0.65}{$L$}} \gamma_\mu b_{\scalebox{0.65}{$L$}} )  
\scalebox{1.0}{$\sum\nolimits_\ell$} ( \bar{\ell} \gamma^\mu \gamma_5 
\ell ) \, , 
\end{array}
\eeq
where the sum over $\ell$ extends over all lepton fields, $e$ ($\gs$)
is the electromagnetic (strong) coupling constant, $q_L$ and $q_R$ are
the chiral quark fields, $F_{\mu \nu}$ ($G_{\mu \nu}^a$) is the
electromagnetic (gluonic) field strength tensor, and $T^a$ are the
color matrices, normalized so that $\mbox{Tr} (T^a T^b) =
\delta^{ab}/2$.  

Unlike the case of $\btosgamma$, the $\btoslpluslminus$ amplitude
involves large logarithms even in the absence of QCD interactions, 
since the current-current operators $Q_1$ and $Q_2$, as well as
$Q_3$--$Q_6$ mix into the vector-like semileptonic operator $Q_9$ at
the one-loop level. QCD contributions from the magnetic operators
$Q_7^\gamma$ and $Q_8^g$, and the axial-vector-like semileptonic
operator $Q_{10}$ enter formally at the NLO, but turn out to be
numerically an $\ord (1)$ correction to the LO result. In order to 
gain an accuracy below the $10 \%$ level on the $\btoslpluslminus$
decay rate a complete resummation of NNLO QCD logarithms has thus to
be performed. Contrary thereto, a similar precision is achieved for
$\btosgamma$ already at the
NLO,\tinyskip\cite{Gambino:2001ew,gorbahnandneubert} accentuating once
again the difference between rare and radiative modes in view of the
RG improved perturbation theory.       

\section{Recent Perturbative Standard Model Calculations}  
\label{sec:calculations}

The aforementioned NNLO QCD computation has required the computation
of $i )$ the $\ord (\as)$ corrections to the relevant Wilson
coefficients,\tinyskip\cite{Bobeth:1999mk} $ii )$ the $\ord(\as)$
contributions to the associated matrix
elements,\tinyskip\cite{Ghinculov:2003qd,matrixelements,Ghinculov:2002pe,Asatryan:2002iy,Asatrian:2003yk,Bobeth:2003at}
and $iii )$ the $\ord (\as^2)$ Anomalous Dimension Matrix (ADM)
describing the mixing of physical dimension-five and six
operators.\tinyskip\cite{oldadm,newadm} Nearly all the ingredients of
the NNLO QCD calculation involve a considerable degree of technical
sophistication and have been performed independently by at least two
groups, sometimes using different methods. However, the most complex
part of the whole enterprise, the calculation of the three-loop $\ord
(\as^2)$ ADM describing the mixing of $Q_1$--$Q_6$ into $Q_1$--$Q_9$
has been completed only very recently.\tinyskip\cite{newadm} Although the
$\ord (\as^2)$ matrix elements of $Q_3$--$Q_6$ have not been
calculated so far,\tinyskip\footnote{In $\BtoXsgamma$ the $\ord
(\as^2)$ matrix elements of $Q_3$--$Q_6$ reduce the branching ratio
by around $1 \%$.\tinyskip\cite{qcdpenguinmatrixelements}} the NNLO
QCD computation of $\btoslpluslminus$ can be called practically
complete, as the $\ord (\as^2)$ matrix elements of $Q_1$ and
$Q_2$,\tinyskip\cite{Ghinculov:2003qd} as well as 
$Q_9$,\tinyskip\cite{Bobeth:2003at} evaluated for arbitrary $q^2$, are
now also available. In their sum the numerical impact of the NNLO QCD
corrections on the differential decay rate amounts to around $-20 \%$
($-25 \%$) in the low-$q^2$ (high-$q^2$) region, and leads to a
reduction of the renormalization scale uncertainties from around $\pm
20 \%$ ($\pm 15 \%$) to $\pm 5 \%$ ($\pm 3 \%$). In the case of the FB
asymmetry, NNLO QCD corrections are not less important, as they shift
$q_0^2$ by around $+15 \%$ and reduce the renormalization scale
dependences from around $\pm 20 \%$ to $\pm 3 \%$.  

In contrast to the track record of QCD corrections, the possible
importance of subleading electroweak effects in $\btoslpluslminus$ has
been realized only quite recently. As shown in the case of radiative
decays,\tinyskip\cite{radiativeelectroweak} they may be as important
as the higher order QCD effects. $\BRll$ is generally parameterized in
terms of $\aem$, but the scale at which $\aem$ should be evaluated is,
in principle, undetermined until higher order electroweak effects are 
taken into account. This has led most authors to use $\aem (\mll)
\approx \aem (\mb) \approx 1/133$. Indeed, in the absence of an $\ord
(\aem)$ calculation, there is no reason to consider $\aem(\mll)$ more
appropriate than, say, $\aem (\MW) \approx 1/128$. As $\BRll$ is
proportional to $\aem^2$, the ensuing uncertainty of almost $8 \%$  
is, compared to the precision achieved in the QCD calculation, not at 
all negligible.\tinyskip\cite{Bobeth:2003at,Chankowski:2003wz} This
uncertainty, which has affected previous analyses, has been recently  
reduced to $2 \%$, by a calculation of the numerical dominant
subleading electroweak effects,\tinyskip\cite{Bobeth:2003at} showing
that after the inclusion of the latter corrections the appropriate
prefactor in $\BRll$ is in fact $\aem (\mb)$. Due to accidental
cancellation, the impact of the calculated electroweak effects is
however rather limited: it amounts to almost $-2 \%$ in $\BRll$,
whereas $q_0^2$ is changed by around $+ 2\%$, and leaves the scale
uncertainties practically unchanged.   

\section{Phenomenology}  
\label{sec:phenomenology}
 
In order to cancel the strong $\mb$-dependence of the differential
$\BtoXslpluslminus$ decay rate, it is customary to normalize the
latter to the experimental value of the Branching Ratio (BR) for the
inclusive  semileptonic decay $\BR[\BtoXclnu]$. However, this
normalization introduces a strong $\mc$-dependence, which is not known
very accurately. The ensuing uncertainty is about $8
\%$.\tinyskip\cite{Ghinculov:2003qd} An alternative
procedure\tinyskip\cite{Gambino:2001ew,Chankowski:2003wz} consists in
normalizing the $\BtoXslpluslminus$ decay width to
$\Gamma[\BtoXulnu]$, and then to express $\BR[\BtoXulnu]$ in terms of
$\BR[\BtoXclnu]$ and of the ratio 
\beq \label{eq:nonperturbativephasespacefactor} 
C = \left | \f{V_{ub}}{V_{cb}} \right|^2
\f{\Gamma[\BtoXclnu]}{\Gamma[\BtoXulnu]} \, ,   
\eeq
which can be computed with better
accuracy.\tinyskip\cite{Gambino:2001ew} In other words, defining $\hat
s \equiv q^2/\mb^2$, one can write the normalized differential decay
rate as    
\beq \label{eq:normalizeddifferentialrate} 
R (\hat s) = \f{\BR[\BtoXclnu]}{C} \left | \f{V_{ub}}{V_{cb}}
\right|^2 \, \f{1}{\Gamma[\BtoXulnu]} \f{d
\Gamma[\BtoXslpluslminus]}{d \hat s} \, ,    
\eeq
where
\bea \label{eq:differentialrate}
& & \hspace{-1.0cm} \f{d \Gamma[\BtoXslpluslminus]}{d {\hat s}} =
\f{\GF^2 m_{b, {\rm pole}}^5 \left | V_{ts}^* V_{tb} \right |^2}{48
\pi^3} \left ( \f{\aem (\mb)}{4 \pi}\right )^2 (1 - {\hat s})^2 \,
\Bigg \{ \left (  4 + \f{8}{{\hat s}} \right ) \Big | 
\widetilde{C}_{7,R}^{\rm eff}  (\hat s) \Big |^2 \non \\   
& & \hspace{-0.75cm} + (1 + 2 {\hat s}) \left ( \Big |
\widetilde{C}_{9, R}^{\rm eff} ({\hat s}) \Big |^2+ \Big |
\widetilde{C}_{10, R}^{\rm eff}({\hat s}) \Big |^2 \right) + 12 {\, \rm 
Re} \left ( \widetilde{C}_{7, R}^{\rm eff} (\hat s)
\widetilde{C}_{9, R}^{\rm eff} (\hat s)^* \right )+ \f{d
\Gamma^{\rm Brems}}{d\hat s} \Bigg \} \, ,  
\eea
and the effective Wilson coefficients carrying the label $R$, include
all real and virtual $\ord (\as)$
corrections,\tinyskip\cite{Ghinculov:2003qd,matrixelements,Ghinculov:2002pe}
whereas the last term denotes the finite $\ord (\as)$ bremsstrahlungs 
corrections.\tinyskip\cite{Asatryan:2002iy}     

\begin{figure}[t]
\begin{center}
\hspace{5mm}
\scalebox{0.425}{\includegraphics{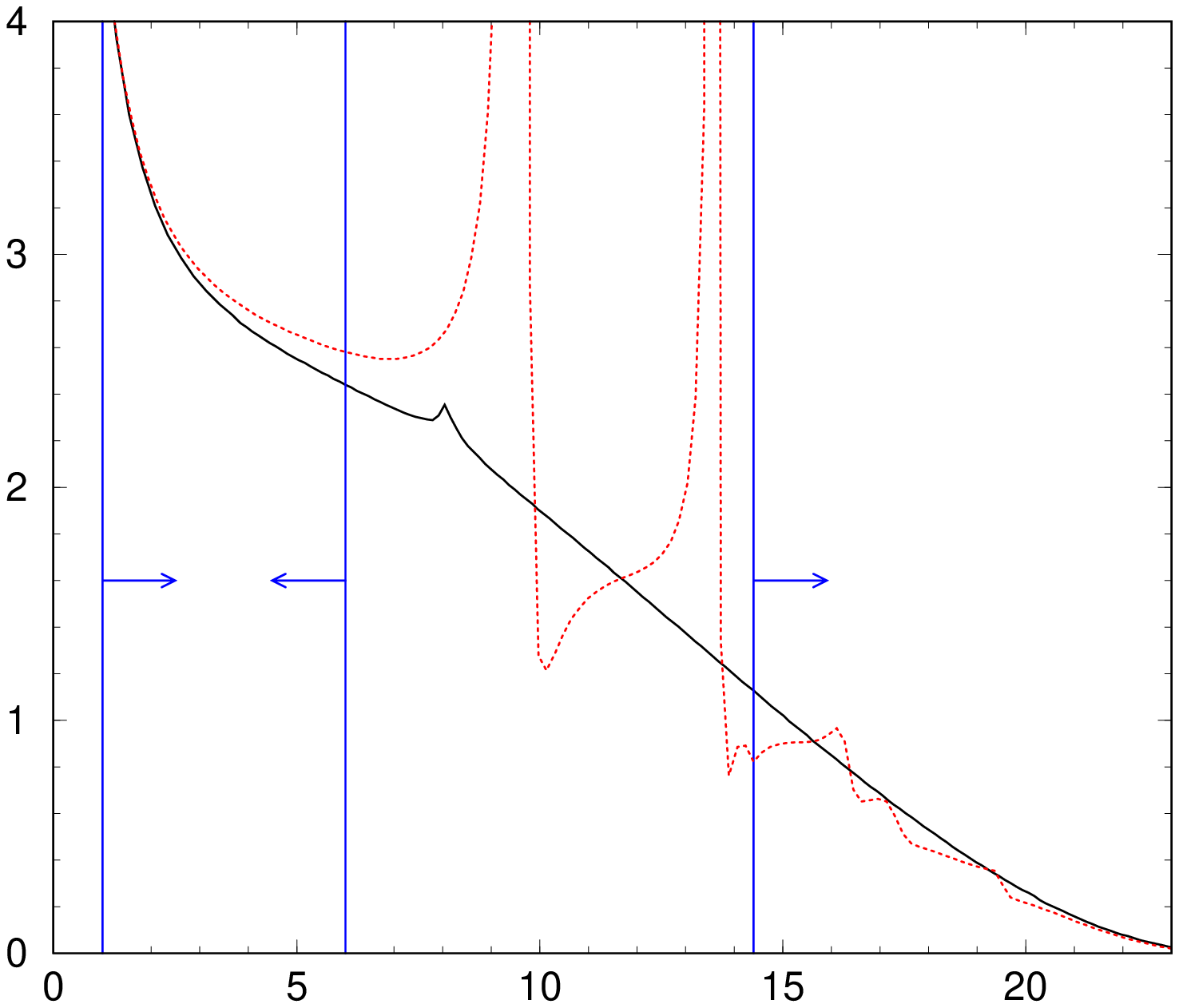}}
\hspace{1.5cm}
\scalebox{0.425}{\includegraphics{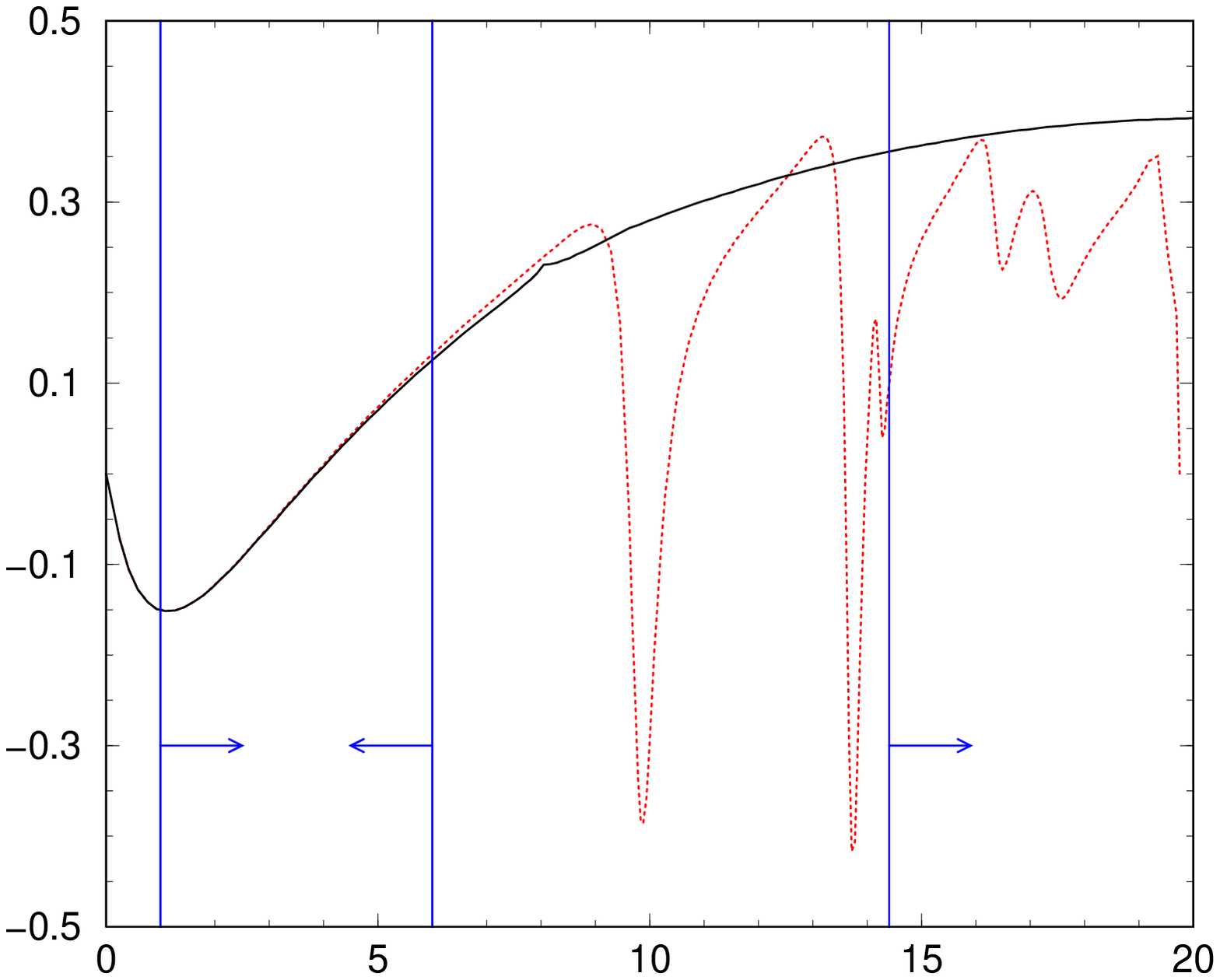}}
\end{center}

\begin{picture}(0,0)%
\setlength{\unitlength}{1pt}%
\put(10,58){{\raisebox{0mm}[0mm][0mm]{%
           \makebox[0mm]{\scalebox{0.75}{\rotatebox{90}{$\f{d \BRll}{d
q^2} \times 10^7~ \left [ \GeV^{-2} \right ]$}}}}}}%
\put(115,0){{\raisebox{0mm}[0mm][0mm]{%
           \makebox[0mm]{\scalebox{0.75}{$q^2~ \left [ \GeV^2 \right
]$}}}}}%
\put(245,80){{\raisebox{0mm}[0mm][0mm]{%
           \makebox[0mm]{\scalebox{0.75}{\rotatebox{90}{$\bar{A}_{\rm
FB} \left ( q^2 \right )$}}}}}}%
\put(360,0){{\raisebox{0mm}[0mm][0mm]{%
           \makebox[0mm]{\scalebox{0.75}{$q^2~ \left [ \GeV^2 \right
]$}}}}}%
\end{picture}%

\caption{NNLO QCD predictions of $d \BRll/d q^2$ (left) and of
$\bar{A}_{\rm FB} \left ( q^2 \right )$ (right) with (dotted red line)
and without (solid black line) a guesstimate of $c \bar c$
effects. The solid blue lines indicate the boundaries of the reference
regions.}        
\label{fig:drfba}
\end{figure}

The use of the $b$-quark pole mass in \Eq{eq:differentialrate} and in
the calculation of the semileptonic width leads to large perturbative
QCD corrections both in the numerator and denominator of
\Eq{eq:normalizeddifferentialrate}. Since this is an artifact of the 
choice of scheme for the $b$-quark mass, they clearly tend to cancel
in the ratio. As a second improvement with respect to previous 
analyses, we  keep terms through $\ord (\as^2)$ in the denominator
and expand the ratio in \Eq{eq:normalizeddifferentialrate} in powers
of $\as$. By making explicit the cancellation of large contributions,
the convergence and stability of the perturbative series improves, 
as we have verified explicitly. Like in previous analyses, due to the
peculiarity of the perturbative expansion for $\btoslpluslminus$, we
retain some large scheme-independent higher order terms in the
amplitude squared: in particular, all terms in
\Eq{eq:differentialrate} that are quadratic in the effective Wilson 
coefficients are expanded in $\as$ and terms up to $\ord(\as)$ are
kept. A careful study shows,\tinyskip\cite{Bobeth:2003at} that with
the two aforementioned improvements the residual scale dependence is
about $50 \%$ smaller than in all preciding analyses.  

After including power corrections of
$\ord(1/\mb^2)$\Tinyskip\cite{mbpowercorrections} and
$\ord(1/\mc^2)$\Tinyskip\cite{Buchalla:1997ky} 
to the differential decay rate of $\BtoXslpluslminus$, as well as 
$\ord(1/\mb^2)$ corrections to $\Gamma[\BtoXclnu]$ and
$\Gamma[\BtoXulnu]$, and expanding \Eq{eq:differentialrate} in
inverse powers of $\mb$ and $\mc$, we obtain for the normalized
differential rate integrated over the low-$q^2$ region      
\beq \label{eq:lowbranchingratio}
\BRll \left ( 1 \GeV^2 \le q^2 \le 6 \GeV^2 \right ) =  \left ( 
1.57^{+0.11}_{-0.10} \big |_{\mt} {^{+0.07}_{-0.07}} \big |_{\mb}
{^{+0.06}_{-0.07}} \big |_{\rm scale} {^{+0.05}_{-0.05}} \big |_{c
\bar{c}} {^{+0.05}_{-0.05}} \big |_{C} \right ) \times 10^{-6} \, . 
\eeq 
Including other subleading parametric uncertainties, the total error 
is about $10 \%$ and is dominated by the uncertainty on the $t$-quark
mass, which will soon be reduced by a factor of two  
by CDF and D0. Moreover, the substantial error from the $b$-quark mass
is an artifact of the employed scheme, which could be reduced
drastically by changing the latter. Hence it should not be interpreted
as a limitation.   

Similarly, we find for the high-$q^2$ window
\beq \label{eq:highbranchingratio}
\BRll \left ( q^2 \ge 14.4 \GeV^2 \right ) =  \left (  
4.02^{+0.71}_{-0.71} \big |_{\mb} {^{+0.24}_{-0.23}} \big |_{\mt}
{^{+0.13}_{-0.13}} \big |_{\rm scale} {^{+0.13}_{-0.13}} \big |_{c
\bar{c}} {^{+0.12}_{-0.12}} \big |_{C} \right ) \times 10^{-7} \, ,  
\eeq 
which is dominated by an error of around $15 \%$ related to the
kinematical cut on $q^2$.\tinyskip\cite{Ghinculov:2003qd} In fact,
even though \Eq{eq:differentialrate} is not explicitly affected by
$\ord (1/\mb)$ corrections, the physical observable is sensitive to
$\ord (1/\mb)$ terms, that arise from the well-known relation between
the mass of the $B$-meson and the $b$-quark in heavy quark effective
theory.    

Finally, our prediction for the integral of the partonic decay rate over the
full spectrum reads  
\beq \label{eq:branchingratiotheory}
\BRll \left ( q^2 \ge 4 m_\mu^2 \right ) =  \left ( 4.58 \pm 0.18_{\rm
scale} \pm 0.66_{\rm para} \right ) \times 10^{-6} \, , 
\eeq
where the parametric error takes into account, besides the power
corrections of $\ord(1/\mb^2)$ and $\ord(1/\mc^2)$, a guesstimate of
the uncertainty related to higher $c \bar c$ resonances, evaluated by 
means of experimental data on $e^+ e^- \to X_c$ using a dispersion
relation.\tinyskip\cite{Kruger:1996cv} The numerical size of the
latter correction can be easily assessed from the left plot of
\Fig{fig:drfba},\tinyskip\footnote{We are grateful to G.~Isidori for
providing us with the two plots shown in \Fig{fig:drfba}.} showing the
NNLO QCD prediction for the differential branching ratio as a function
of $q^2$.       

Unfortunately for all of us who hoped to discover NP in rare $\btos$
transitions, the estimate in \Eq{eq:branchingratiotheory} compares
fairly good with the recent experimental world
average\tinyskip\cite{Nakao:2003gc}   
\beq \label{eq:branchingratioexperiment} 
\BRllexp = \left ( 6.2^{+1.1}_{-1.1} \big |_{\rm stat} {^{+1.6}_{-1.3}
\big |_{\rm syst} } \right ) \times 10^{-6} \, .     
\eeq
Even worse, it agrees amazingly well with the preliminary result by
Belle,\tinyskip\cite{iwasaki} which suggests a somewhat lower central
value for the combination of all inclusive measurements.  

Facing such bad news, lets turn our attention to the FB asymmetry,
which in its so-called normalized form is defined as  
\beq \label{eq:forwardbackwardasymmetry}
\bar{A}_{\rm FB}(\hat s)=\f{1}{d \Gamma[\BtoXslpluslminus]/d \hat{s}}
\int_{-1}^1 \! d\cos\theta_\ell \, \f{d^2
\Gamma[\BtoXslpluslminus]}{d\hat s \,d\cos\theta_\ell} \,{\rm
sgn}(\cos\theta_\ell) \, , 
\eeq
where 
\bea \label{eq:doubledifferentialrate}
& & \hspace{-1cm} \int_{-1}^1 \! d\cos\theta_\ell \, \f{d^2
\Gamma[\BtoXslpluslminus]}{d\hat s \,d\cos\theta_\ell} \,{\rm 
sgn}(\cos\theta_\ell) = \f{\GF^2 m_{b, {\rm pole}}^5 \left | V_{ts}^*
V_{tb} \right |^2}{48 \pi^3} \left ( \f{\aem (\mb)}{4 \pi}\right )^2
(1 - {\hat s})^2 \non \\ 
& & \hspace{-0.75cm} \times \Bigg \{ \! -6 \, {\rm Re} \left ( 
\widetilde{C}^{\rm eff}_{7, {\rm FB}} (\hat s) \widetilde{C}^{{\rm
eff}}_{10, {\rm FB}} (\hat s)^\ast \right ) - 3 \hspace{0.1mm} \hat{s}
\, {\rm Re} \left ( \widetilde{C}^{\rm eff}_{9, {\rm FB}} (\hat s)
\widetilde{C}^{{\rm eff}}_{10, {\rm FB}} (\hat s)^\ast \right ) +
A_{\rm FB}^{\rm Brems} (\hat s) \Bigg \} \, , 
\eea
and $\theta_\ell$ denotes the angle between the momentum of the
positively charged lepton and the $B$-meson in the rest frame of the
lepton pair. The effective Wilson coefficients in
\Eq{eq:forwardbackwardasymmetry} take into account all $\ord (\as)$
real and virtual
corrections,\tinyskip\cite{Ghinculov:2003qd,matrixelements,Ghinculov:2002pe}
as indicated by the subscript $\rm FB$, whereas the last term encodes
the $\ord (\as)$ bremsstrahlung
corrections,\tinyskip\cite{Asatrian:2003yk} which have not been 
included in our analysis of $q_0^2$, since they turn out to be below
$1 \%$. The NNLO QCD prediction of the normalized FB asymmetry as a
function of $q^2$ is presented in the right plot of \Fig{fig:drfba},
which also illustrates the numerical size of the long-distance
contributions due to intermediate $c \bar c$ resonances. 

Since $q_0^2$ is known to be especially sensitive to physics beyond
the SM, a comprehensive study of the residual theoretical error
attached to it is important. As pointed out
earlier,\tinyskip\cite{Ghinculov:2002pe} the renormalization scale 
dependence of $q_0^2$ computed from \Eq{eq:doubledifferentialrate} is
rather small. However, an alternative way to estimate the residual 
theoretical error follows from the observation that $q_0^2$ should be
independent of the normalization for the FB asymmetry. As both
numerator and denominator of \Eq{eq:forwardbackwardasymmetry} are
truncated series in $\as$, one can expand
\Eq{eq:forwardbackwardasymmetry} in $\as$, making $q_0^2$ 
sensitive to different combinations of higher order terms, that depend
on the adopted normalization. A thorough analysis
shows,\tinyskip\cite{Bobeth:2003at} that the dependence of $q_0^2$ on 
the specific method used to compute it is generally larger than the
scale uncertainty associated with it. Implementing all known
perturbative and non-perturbative corrections, and assigning to the
central value a rather conservative theoretical error of $6 \%$, that
covers the whole range of possible values obtained by changing the
normalization in \Eq{eq:forwardbackwardasymmetry}, we finally get   
\beq \label{eq:zeroforwardbackward}
q^2_0 = \left ( 3.76  \pm {0.22}_{\rm theory} \pm 0.24_{\mb} \right )
\GeV^2 \, ,      
\eeq
with a total error close to $9 \%$. Also in this case the error due to
the $b$-quark mass can be drastically reduced by performing the
calculation in a different mass scheme.   

\section{Outlook}
\label{sec:outlook}

Rare semileptonic $B$ decays are going to play an important role in 
the search for NP at the $B$-factories and upcoming flavor
physics experiments at the Tevatron and the LHC. From the theoretical
point of view the inclusive mode in both the low-$q^2$ and high-$q^2$ 
window is particular interesting, since it can be accurately computed
in the SM. Given this situation, a measurement of the differential
decay rate integrated separately over the two reference regions  
would be desirable. Furthermore, as the FB asymmetry can change
drastically in scenarios of NP, even a rather crude measurement of its
shape would either rule out large parts of the parameter space of the
underlying model or show clear evidence for physics beyond the SM. 

\section*{Acknowledgments}

First of all, let us apologize for the cover picture of this talk,
which shows the author only rather vaguely. We are grateful to
C.~Bobeth, M.~Gorbahn and P.~Gambino for fruitful collaboration,
G.~Isidori for useful communication, and G.~Zanderighi for a careful
proofreading of the manuscript. Before drawing the curtain, we finally
thank M.~Carena and C.~E.~M.~Wagner for the permission to display the
photo of their son Julian, taken by L.~Fayard. This work is supported
by the U.S. Department of Energy under contract No. DE-AC02-76CH03000.

\end{document}